\documentclass{elsart}

\usepackage{epsfig}
\usepackage{amssymb}

\begin{document}

\begin{frontmatter}

\title {Temperature dependence of the upper critical field of 
high-$T_{c}$ superconductors from isothermal magnetization data. 
Influence of a temperature dependent Ginzburg-Landau parameter.}

\author[label1,label2]{I. L. Landau}
\author[label1]{H. R. Ott}
 \address[label1]{Laboratorium f\"{u}r Festk\"{o}rperphysik, 
 ETH-H\"{o}nggerberg, CH-8093 Z\"{u}rich, Switzerland}
 \address[label2]{Institute for Physical Problems, 117334 Moscow, Russia}

\begin{abstract}
We show that the scaling procedure, recently proposed for the evaluation 
of the temperature variation of the normalized upper critical field of 
type-II superconductors, may easily be modified in order to take into 
account a possible temperature dependence of the Ginzburg-Landau 
parameter $\kappa$. As an example we consider $\kappa (T)$ as it follows 
from the microscopic theory of superconductivity.
\end{abstract}

\end{frontmatter}

\section {Introduction}

As we have recently shown in Ref.  \cite{1}, the temperature variation 
of the normalized upper critical field $H_{c2}$ of a type-II 
superconductor may be obtained by scaling the data of the reversible 
isothermal magnetization $M$, measured as a function of external magnetic 
fields $H$ at different temperatures. One of the advantages of this 
approach is that no particular \textit{a priori} assumption for the field 
dependence of the magnetization, $M(H)$, needs to be made. This is why 
the proposed scaling procedure is expected to be equally valid for any 
type-II superconductor, regardless of its anisotropy, the type of pairing 
or the value of the Ginzburg-Landau (GL) parameter $\kappa$. The scaling 
relation introduced in Ref.  \cite{1} is based, however, on the assumption 
that $\kappa$ is temperature independent. 
In this particular case, the magnetic susceptibility $\chi (H,T)$ of a 
type-II superconductor in the mixed state is a universal function of 
$H/H_{c2}(T)$ \cite{2} and, according to Ref. \cite{1}, the 
relation between the values of $M$ at two different temperatures is
\begin{equation}
M(H/h_{c2},T_{0}) = M(H,T)/h_{c2} + c_{0}(T)H,
\end{equation}
with $h_{c2} = H_{c2}(T)/H_{c2}(T_{0})$ representing the upper critical 
field, normalized by its value at some chosen temperature $T_{0} < T_{c}$. 
The term $c_{0}(T)H$ of this equation was introduced in order to take into 
account the contribution to the sample magnetization due to the 
temperature dependent paramagnetic susceptibility $\chi_{n}$ of the 
material in the normal state. Since $c_{0}(T) = [\chi_{n}(T) - 
\chi_{n}(T_{0})]$, this second term in Eq. (1) can be omitted, if 
$\chi_{n}$ does not vary with temperature. 

The scaling procedure based on Eq.  (1) was applied to numerous data 
sets of the reversible magnetization of HTSC's which are available in 
the literature and it turned out that it works rather well for single 
crystals and grain aligned samples \cite{1,new}, as well as for HTSC 
ceramics \cite{3}. A surprising result of the work presented in Refs.  
\cite{1,new,3} is that all the different superconducting cuprates may be 
divided into two groups. For each group, the corresponding 
$h_{c2}(T/T_{c})$ curves are practically identical for all materials of 
this group.  The smaller group of the two appears to include compounds 
of the series Y$_{2}$Ba$_{4}$Cu$_{7}$O$_{15+x}$ (Y-247), underdoped 
YBa$_{2}$Cu$_{3}$O$_{7-x}$ (Y-123), and 
Tl$_{2}$Ba$_{2}$CaCu$_{2}$O$_{8+x}$ (Tl-2212)  \footnote{The situation 
with this particular Tl compound is not yet clear. While the 
$h_{c2}(T/T_{c})$ curves for a single crystalline sample of Tl-2212 is 
very close to those for Y-247 and underdoped Y-123 compounds 
\cite{new}, a similar curve for a ceramic sample exhibits a quite 
different shape (see the data for sample Tl\#3 in Ref. \cite{3}). It is 
possible that some difference in the oxygen content of these two samples 
is responsible for this effect.} \cite{new,3}. All other varieties of 
HTSC's belong to the other group. This universality of $h_{c2}(T/T_{c})$ 
for seemingly very different HTSC compounds emerges as a rather unexpected 
result but at the same time it indicates that the scaling procedure 
outlined in Ref. \cite{1} is a valid tool for the analysis of data of the 
reversible magnetization in the mixed state of type-II superconductors. 

As was already argued in Ref. \cite{1}, this universality does not 
necessarily mean that the GL parameter $\kappa$ in HTSC's is indeed 
temperature independent. It only indicates that, if $\kappa$ is 
temperature dependent, its temperature dependence is nearly the same for 
all compounds in each of the above mentioned groups. For completeness, 
however, it seems of interest to also check the influence of possible 
temperature variations of $\kappa$ on the resulting $h_{c2}(T/T_{c})$ 
curves and in this work we present and discuss the necessary corrections 
to Eq. (1).  Since a reasonably simple modification of Eq.  (1) can only 
be made if $\kappa \gg 1$, we shall focus our discussion on this 
situation, which is definitely met for cuprate superconductors.

In magnetic fields $H$ much larger than the lower critical field 
$H_{c1}$, the magnetic moment of a superconductor is inversely 
proportional to $\kappa ^{2}$ \cite{2}. This allows for a rather simple 
modification of Eq. (1). Indeed, if $\kappa$ is temperature dependent, 
Eq. (1) is to be be replaced by 
\begin{equation}
M(H/h_{c2},T_0)=\frac{M(H,T)}{[\kappa (T)/\kappa (T_0)]^2 h_{c2}} + 
c_0(T)H.
\end{equation}
The condition $H \gg H_{c1}$ is always satisfied for our type of 
experiments because equilibrium-magnetization data are only accessible 
above the irreversibility field $H_{irr}$. The irreversibility 
line $H_{irr}(T)$ in the $H$-$T$ phase diagram of HTSC's is known to 
correspond to fields much higher than $H_{c1}(T)$. 
\begin{figure}[!t]
 \begin{center}
  \epsfxsize=0.8\columnwidth \epsfbox {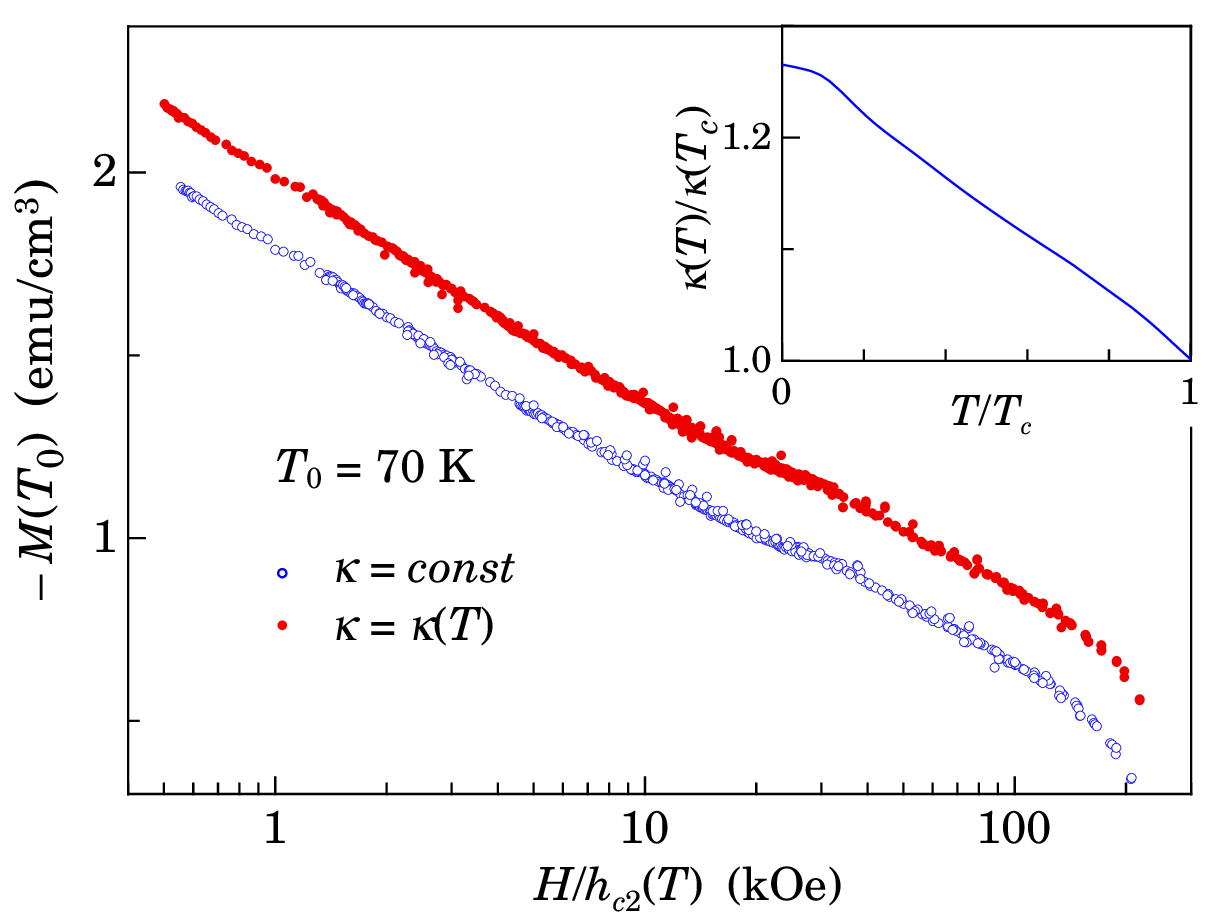}
  \caption{$M_{eff}(H,T_{0})$ calculated using Eq (1) ($\kappa = const.$) 
           and Eq. (2) ($\kappa = \kappa(T)$) for sample Bi\#1. The curve 
           corresponding to $\kappa = \kappa(T)$ is shifted upwards by 
           0.2 emu/cm$^{3}$ for clarity. The inset shows the 
           calculated temperature 
           dependence of the GL parameter $\kappa$ normalized by its value 
           at $T = T_{c}$ (see to Refs. \cite{4} and \cite{5}).}
 \end{center}
\end{figure}

As may be seen, Eq. (2), in comparison with Eq. (1), contains an 
additional unknown parameter $\kappa (T)/ \kappa (T_{0})$. In 
principle, all three parameters $h_{c2}(T)$, $c_{0}(T)$ and $\kappa 
(T)/ \kappa (T_{0})$ may be evaluated by scaling the magnetization data 
collected at different temperatures with the procedure which is 
described in Ref. \cite{1}. However, because reversible $M(H)$ data for 
HTSC's are only available in a rather limited range of magnetic fields 
and because two of the unknown parameters appear together as a product in 
the denominator of the first term on the right hand side of Eq. (2), 
this scaling procedure cannot deliver all three parameters 
independently with 
adequate certainty. Thus, in order to use Eq. (2) for the scaling 
procedure, the temperature dependence of $\kappa(T)$ must, as before, 
{\it a priori} be assumed.\footnote{The previous assumption, in Ref. 1, 
was $\kappa = const$} In the following discussion we consider, as 
an example, $\kappa(T)$ as it is calculated from the microscopic theory 
of superconductivity \cite{4,5} (see the inset of Fig. 1). Once this 
assumption is made, Eq.  (2) may be used for the analysis of 
experimental data exactly in the same way as Eq. (1) was employed in
our previous studies \cite{1,new,3}. In the following we denote the 
values of $M$ calculated from Eqs. (1) or (2) as $M_{eff}(H,T_{0})$, 
simply to distinguish them from experimentally measured magnetizations.

\section {Analysis of experimental data}

We now apply Eq.  (2) for the analysis of magnetization
data that were presented in previous publications of other authors. 
Most of these results were already analyzed with the assumption 
that $\kappa$ does not vary with temperature \cite{1,3}. Some 
characteristics of the samples are listed in Table 1.
\begin{figure}[t]
 \begin{center}
  \epsfxsize=0.8\columnwidth \epsfbox {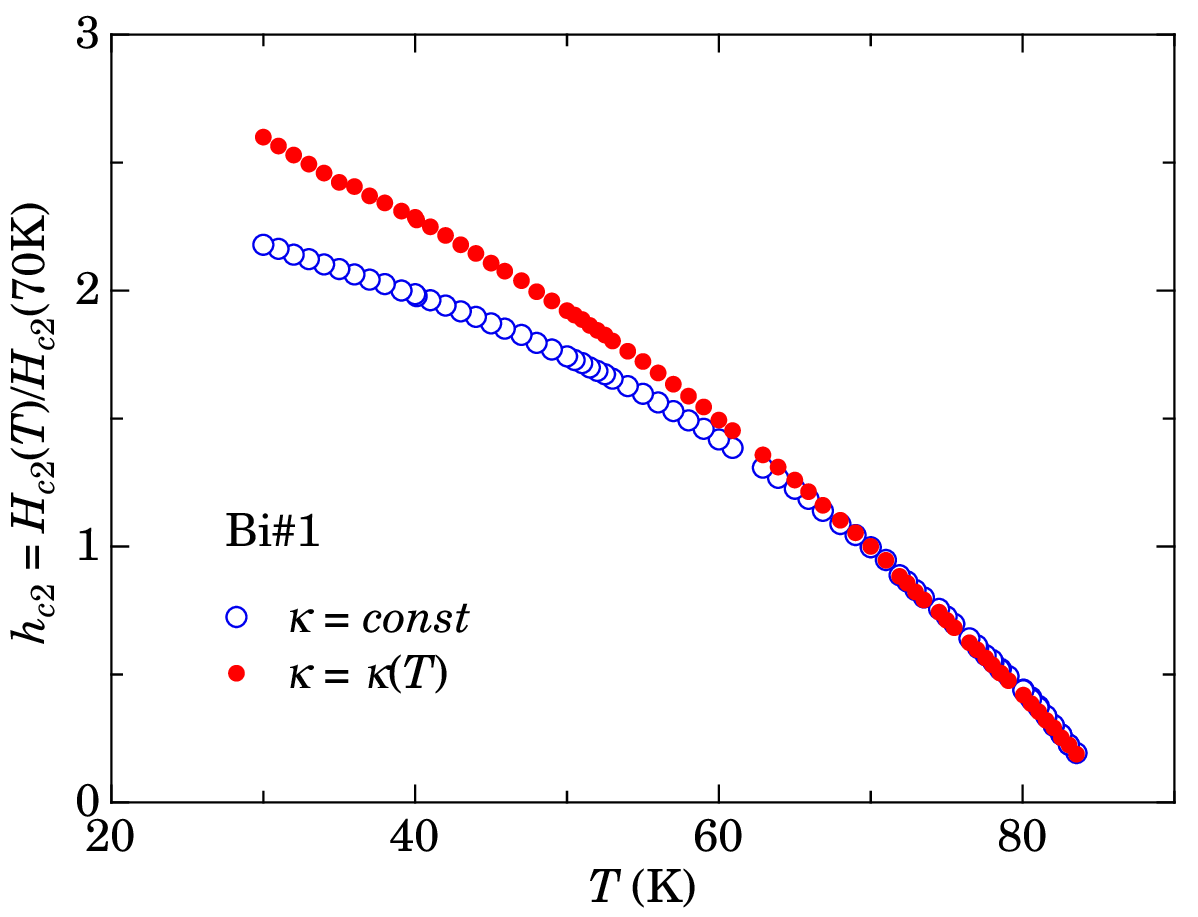}
  \caption{Temperature dependence of $H_{c2}$ normalized by its value 
           at $T_{0} = 70$ K for sample Bi\#1. The curve calculated assuming 
           that the GL parameter is temperature independent ($\kappa = 
           const.$) is shown for comparison.}
 \end{center}
\end{figure}

Eq. (2) may be used to merge the $M(H)$ curves, measured at different 
temperatures, into a single $M_{eff}(H,T_{0})$ curve by a suitable 
choice of scaling parameters $h_{c2}(T)$ and $c_{0}(T)$ (see Ref. 
\cite{1} for details). Fig. 1 displays the resulting $M_{eff}(H,T_{0})$ 
curves calculated using Eq. (1) ($\kappa  = const$) and Eq. (2)  
($\kappa = \kappa (T)$) for sample Bi\#1 (see Table I). The 
$M_{eff}(H,T_{0})$ curves presented in Fig. 1 were obtained from more 
than 60 individual $M(H)$ curves measured at different temperatures 
between 30 and 83.5 K (see Ref. \cite {6} for the original experimental 
data). As may be seen, the shapes of the $M_{eff}(H,T_{0})$ curves, 
representing the field dependence of the equilibrium magnetization at 
$T = T_{0}$, are practically identical and not altered significantly by 
the different assumptions for $\kappa(T)$. 

The resulting temperature dependencies of the normalized upper critical 
fields $h_{c2}$ for the same sample are shown in Fig. 2. A difference 
between the two considered cases may clearly be seen only 
if the data set extends to temperatures well below $T_{c}$. Fig. 3 
displays $H_{c2}(T)/H_{c2}(0.9T_{c})$ data calculated using Eq. (2) for 
several samples listed in Table I. It is obvious that the assumed 
temperature dependence of $\kappa$ does not change the previously 
established fact that the 
$h_{c2}(T/T_{c})$ curves for different HTSC's are identical.  
\begin{figure}[h]
 \begin{center}
  \epsfxsize=0.8\columnwidth \epsfbox {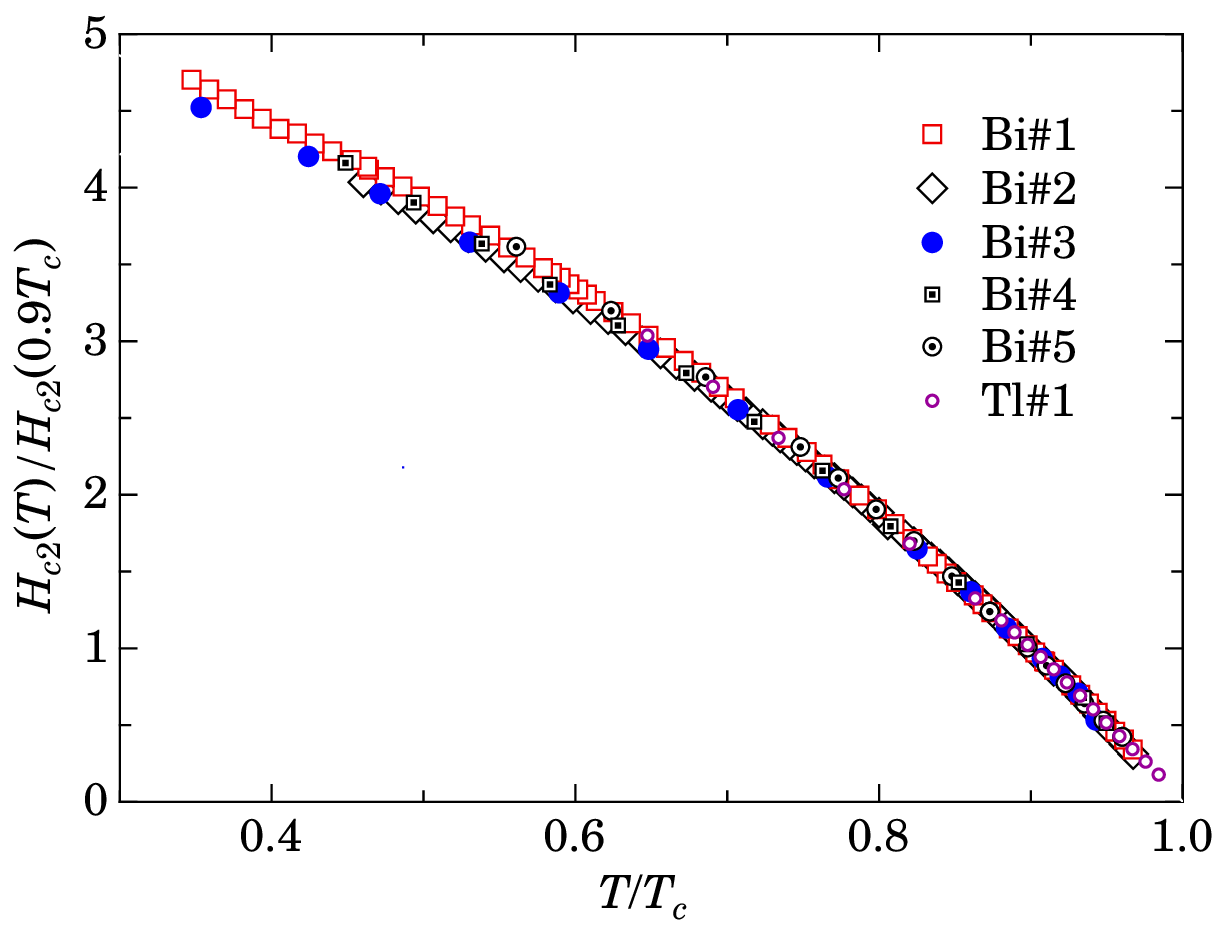}
  \caption{Temperature dependence of $H_{c2}$ normalized by its values 
           at $T = 0.9T_{c}$ for samples listed in Table I.}
 \end{center}
\end{figure}
\begin{figure}[h]
 \begin{center}
  \epsfxsize=0.8\columnwidth \epsfbox {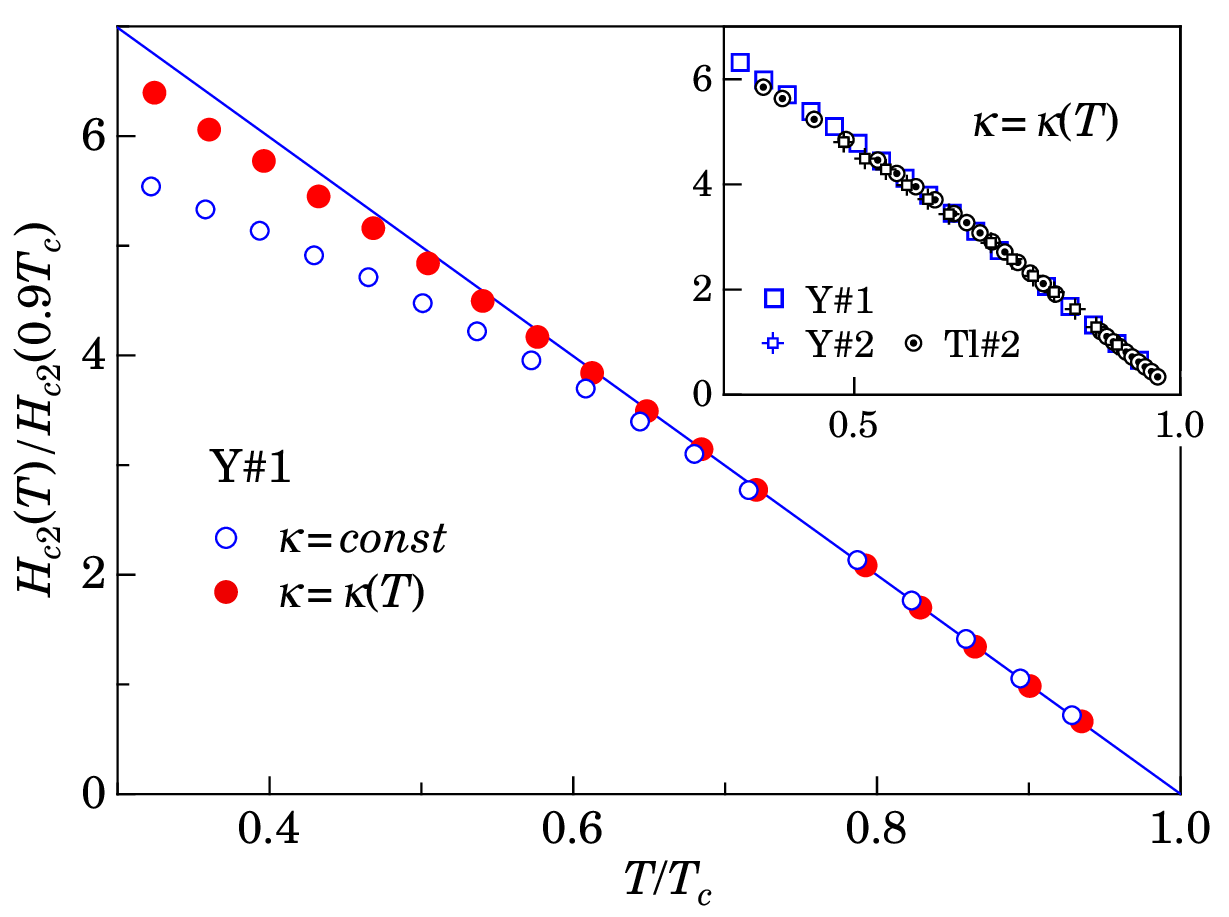}
  \caption{Temperature dependence of $H_{c2}$ normalized by its values 
           at $T = 0.9T_{c}$ for an underdoped Y-based sample Y\# 1. 
           The solid line is the best linear fit to the data points 
           for $T/T_{c} \ge 0.8$. The $H_{c2}(T)/H_{c2}(0.9T_{c})$ 
           curve calculated assuming $\kappa$ being independent of 
           temperature is shown for comparison. The inset shows 
           analogous $h_{c2}(T/T_{c})$ data for several samples listed 
           in Table I.}
 \end{center}
\end{figure}

As has briefly been mentioned  above, the $h_{c2}(T)$ curves for Tl-2212, 
Y-247 and oxygen deficient Y-123 compounds are quite different from those 
for many other HTSC's. The result for an oxygen-deficient sample Y-123 
sample is shown in the main-frame Fig. 4. Similar to the case presented in 
Fig. 2, the assumed temperature dependence of $\kappa$ reduces the 
curvature of the calculated $h_{c2}(T)$ curves in comparison with 
$h_{c2}(T)$ for $\kappa = const$.

\section {Conclusion}

We considered the influence of a possible temperature variation of 
the Ginzburg-Landau parameter $\kappa$ on the results that follow from 
the analysis of equilibrium magnetization data of type-II 
superconductors using the scaling procedure proposed in Ref. \cite{1}. 
We showed  that assuming $\kappa(T)$, as it follows from the microscopic 
theory of superconductivity, results in a slight but noticeable change 
of the calculated temperature dependencies of the normalized upper 
critical fields $h_{c2}(T/T_{c})$ in comparison with those that are 
calculated by assuming $\kappa$ as being independent of temperature. The 
main qualitative conclusions following from our previous studies 
\cite{1,new,3} remain, however, unchanged. All investigated HTSC's may 
be divided in two groups with identical $h_{c2}(T/T_{c})$ curves for all 
the compounds belonging to each group. We also note that the $h_{c2}(T)$ 
curves for both groups are qualitatively the same as for conventional 
superconductors. They are linear in $T$ at temperatures close to $T_{c}$ 
with downward deviations from linearity at lower temperatures.

\begin{table}[!h]
	
\caption{Sample identification.}
\begin{tabular}{lcccccc}
\multicolumn{1}{c}{No.} &
\multicolumn{1}{c}{Refs.} &
\multicolumn{1}{c}{Compound} &
\multicolumn{1}{c}{Sample} &
\multicolumn{1}{c}{$T_{c}$ (K)} \\

Bi\#1 & \cite{6} & Bi$_{2.12}$Sr$_{1.9}$Ca$_{1.2}$Cu$_{1.96}$O$_{8+x}$ & 
single crystal &  86.9  \\
Bi\#2 & \cite{6} & Bi$_{2.12}$Sr$_{1.9}$Ca$_{1.2}$Cu$_{1.96}$O$_{8+x}$ & 
ceramic &  86.4\\
Bi\#3 & \cite{7} & Bi$_{2}$Sr$_{2}$Ca$$Cu$_{2}$O$_{8+x}$ & single crystal &  
84.8 \\
Bi\#4 & \cite{8} & Bi$_{2}$Sr$_{2}$Ca$$Cu$_{2}$O$_{8+x}$ & single crystal &  66.8 \\
Bi\#5 & \cite{9} & Bi$_{2}$Pb$_{0.2}$Sr$_{2}$Ca$$Cu$_{2}$O$_{8}$ & single crystal &  86.7 \\
Tl\#1 & \cite{10} & Tl$_{0.7}$Bi$_{0.2}$Sr$_{1.8}$Ba$_{0.2}$Ca$_{1.9}$Cu$_{3}$O$_{x}$ & 
ceramic &  115.8 \\
Tl\#2 & \cite{last} & Tl$_2$Ba$_2$CaCu$_2$O$_{{8 + x}}$ & single crystal &  102.4 \\
Y\#1 & \cite{11} & YBa$_{2}$Cu$_{3}$O$_{6.69}$ & ceramic &  55.5 \\
Y\#2 & \cite{11} & YBa$_{2}$Cu$_{3}$O$_{6.81}$ & ceramic &  62.0 \\
\end{tabular}

\end{table}

\end{document}